\documentstyle[12pt]{article}
\begin{document}
\begin{center}
{\Large QED theory of the nuclear recoil effect in atoms }\\

\end{center}
\begin{center}
{V.M. Shabaev}
\\
\end{center}

\begin{center}
{\it Department of Physics, St.Petersburg State University,}\\
{\it Oulianovskaya 1, Petrodvorets, St.Petersburg 198904, Russia}
\end{center}

PACS number(s): 12.20.-m, 31.30.Jv, 31.30.Gs
\begin{abstract}
The quantum electrodynamic theory of the nuclear recoil
effect in atoms to all orders in $\alpha Z$ is
formulated. The nuclear recoil corrections for
 atoms with one and two electrons
over closed shells are considered in detail.
The problem of the composite nuclear structure
in the theory of the nuclear recoil effect is discussed.
\end{abstract}
\newpage
\section{Introduction}

The complete $\alpha Z$-dependence expressions for the nuclear
recoil corrections to the energy levels of hydrogenlike atoms
were first derived in [1].
These expressions consist of three contributions: the Coulomb
contribution, the one-transverse-photon contribution, and
the two-transverse-photon contribution.
For a state $a$
the Coulomb contribution is given by
(the relativistic units $\hbar=c=1$ are used in the paper)
\begin{eqnarray}
\Delta E_{c}&=&\Delta E_{c}^{(1)}+\Delta E_{c}^{(2)}\,,\nonumber\\
\Delta E_{c}^{(1)}&=&\langle a|\frac{{\bf p}^{2}}{2M}|a\rangle\,,\\
\Delta E_{c}^{(2)}&=&\frac{2\pi i}{M}\int_{-\infty}^{\infty}d\omega\,
\delta_{+}^{2}(\omega)\langle a|[{\bf p},V_{c}]G(\omega+\varepsilon_{a})
[{\bf p},V_{c}]|a\rangle\,,
\end{eqnarray}
where $|a\rangle$ is the unperturbed state of the Dirac electron
in the Coulomb field of the nucleus,
$V_{c}=-\frac{\alpha Z}{r}$
 is the Coulomb potential of the nucleus,
${\bf p}$ is the momentum operator, $\delta_{+}(\omega)=\frac{i}{2\pi}
(\omega+i0)^{-1}$,
 $G(\omega)=(\omega-H(1-i0))^{-1}$ is the relativistic
Coulomb Green function,
$ H=
\mbox{\boldmath $\alpha$}
{\bf p}+\beta m +V_{c}\,$.
The one-transverse-photon contribution is
\begin{eqnarray}
\Delta E_{tr(1)}&=&\Delta E_{tr(1)}^{(1)}+\Delta E_{tr(1)}^{(2)}\,,
\nonumber\\
\Delta E_{tr(1)}^{(1)}&=&-\frac{1}{2M}\langle a|\Bigl({\bf D}(0){\bf p}+
{\bf p}{\bf D}(0)\Bigr)|a\rangle\,,\\
\Delta E_{tr(1)}^{(2)}&=&-\frac{1}{M}\int_{-\infty}^{\infty}d\omega\,
\delta_{+}(\omega)\langle a|\Bigl([{\bf p},V_{c}]G(\omega+\varepsilon_{a})
{\bf D}(\omega)\nonumber\\
&&-{\bf D}(\omega)G(\omega+\varepsilon_{a})[{\bf p},V_{c}]
\Bigr)|a\rangle\,,
\end{eqnarray}
where
$$
D_{m}(\omega)=-4\pi\alpha Z\alpha_{l}D_{lm}(\omega)\,,
$$
$\alpha_{l}\;(l=1,2,3)$ are the Dirac matrices, $ D_{lm}(\omega)$ is
the transverse part of the photon propagator in the Coulomb gauge.
In the coordinate representation it is
$$
D_{ik}(\omega,{\bf r})=-\frac{1}{4\pi}\Bigl\{\frac
{\exp{(i|\omega|r)}}{r}\delta_{ik}+\nabla_{i}\nabla_{k}
\frac{(\exp{(i|\omega|r)}
-1)}{\omega^{2}r}\Bigr\}\,.
$$
 The two-transverse-photon
contribution is
\begin{eqnarray}
\Delta E_{tr(2)}=\frac{i}{2\pi M}\int_{-\infty}^{\infty}d\omega\,
\langle a|{\bf D}(\omega)G(\omega +\varepsilon_{a}){\bf D}(\omega)|a\rangle\,.
\end{eqnarray}

An attempt to derive the complete $\alpha Z$-dependence
expressions for the nuclear recoil effect was previously
undertaken in [2]. Except for the Coulomb contribution,
the expressions found in [2] are in disagreement with
the ones given above.
A dominant part of this disagreement is caused by technical errors
made in [2]. If we remove these errors from [2],
a discrepancy remains in the one-transverse-photon contribution
and, in addition, appears in the Coulomb contribution. This
discrepancy was discussed in detail in [1].

Recently, the equations (1)-(5) were rederived
in [3,4]. In Ref. [3], it was noted that the sum of these expressions
can be written in the following compact form
\begin{eqnarray}
\Delta E_{tot}=\frac{i}{2\pi M}\int_{-\infty}^{\infty}d\omega\,
\langle a|({\bf p} -{\bf D}(\omega))G(\omega +\varepsilon_{a})
({\bf p} - {\bf D}(\omega))|a\rangle\,.
\end{eqnarray}

The  terms $\Delta E_{c}^{(1)}$ and $\Delta E_{tr(1)}^{(1)}$
 can easily be calculated by using
the virial relations for the Dirac  equation [5]. Such a calculation
 gives [1]
\begin{eqnarray}
\Delta E^{(1)}&\equiv&\Delta E_{c}^{(1)}+\Delta E_{tr(1)}^{(1)}=
\frac{m^{2}-\varepsilon_{a}^{2}}{2M}\,.
\end{eqnarray}
This simple formula contains all the nuclear recoil corrections
within the $(\alpha Z)^{4}m^2/M$ approximation.
The remaining terms (Eqs. (2),(4), and (5)) taken to the
lowest order in $\alpha Z$ give the Salpeter  corrections [6].
Evaluation of these terms to all orders in $\alpha Z$
in the range $Z=1-100$ was done in [7].
In particular, it was found in [7] that the complete (in $\alpha Z$)
 nuclear recoil correction, additional to the Salpeter one,
to the Lamb shift ($n=2$) in hydrogen constitutes -1.32(6) kHz.
This value almost coincides with the value of the
 $(\alpha Z)^{6}m^2/M$ correction found in [4,8-10].

The complete $\alpha Z$-dependence expressions for the nuclear recoil
corrections for high $Z$ few-electron atoms were derived in [11].
These formulas were used in [7] to calculate the nuclear recoil
corrections to all orders in $\alpha Z$ for  high $Z$ lithiumlike
atoms. As it follows from these formulas, within
 the $(\alpha Z)^{4}m^{2}/M$
approximation the nuclear  recoil corrections
 can be obtained
by averaging the operator
\begin{eqnarray}
H_{M}=\frac{1}{2M}\sum_{s,s'}\Bigl({\bf p}_{s}\cdot{\bf p}_{s'}-
\frac{\alpha Z}{r_{s}}\Bigl(\mbox{\boldmath $\alpha$}_{s}+\frac
{(\mbox{\boldmath $\alpha$}_{s}\cdot{\bf r}_{s}){\bf r}_{s}}{r_{s}^{2}}
\Bigr)\cdot{\bf p}_{s'}\Bigr)
\end{eqnarray}
with the Dirac wave functions.
An independent derivation of this operator was done in [12].
The operator (8) was employed in [13] to calculate
the $(\alpha Z)^{4}m^{2}/M$ corrections
to the energy levels of two-
 and three-electron multicharged ions.

In the present paper we generalize the theory
of the nuclear recoil effect to
an arbitrary case of a many-electron atom.
In particular, this generalization allows one to use as the zero
approximation a potential that is different from the
pure Coulomb field. In addition, it allows one to use
the formalism in which the closed shells are refered to
the vacuum state.
 In Sec. 2, we formulate the basic equations
of the method. In Sec. 3 ,  we apply this method to an atom
with one electron over closed shells. In Sec. 4, the case
of an atom with two electrons over closed shells is considered.
In Sec. 5, the problem of the composite
nuclear structure is discussed.

\section{Basic formalism}

Like Refs. [3,4], we will consider the nucleus as a non-relativistic
particle.  In the Schr\"odinger representation and
 the Coulomb gauge, the Hamiltonian of the whole
system is
\begin{eqnarray}
H&=&\int d{\bf x} \psi^{\dag}({\bf x})
[\mbox{\boldmath $\alpha$}\cdot
(-i\nabla_{\bf x}
-e{\bf A}({\bf x}))+\beta m]\psi({\bf x})\nonumber\\
 &&+\frac{e^{2}}{8\pi}\int d{\bf x}d{\bf y}\,\frac{
\rho_{e}({\bf x})
\rho_{e}({\bf y})}{|{\bf x}-{\bf y}|}+\frac{1}{2}
\int d{\bf x}
({\bf {\cal E}}_{t}^{2}({\bf x})+{\bf {\cal H}}^{2}({\bf x}))
\nonumber\\
 &&+\frac{e|e|Z}{4\pi}\int d{\bf x}\,\frac{
\rho_{e}({\bf x})}{|{\bf x}-{\bf X}_{n}|}+\frac{1}{2M}
({\bf P}_{n}-|e|Z{\bf A}({\bf X}_{n}))^{2}
\nonumber\\
&&-\mbox{\boldmath $\mu$}\cdot
{\bf {\cal H}}({\bf X}_{n})\,,
\end{eqnarray}
where $m$ is the electron mass, $M$ is the nucleus mass,
$e$ is the electron charge ($e < 0$), ${\bf X}_{n}$ is
the radius vector of the nucleus,
 ${\bf P}_{n}=
-i\nabla_{{\bf X}_{n}}$,
and $\mbox{\boldmath $\mu$}$ is the magnetic moment
of the nucleus.  The  term
$-\mbox{\boldmath $\mu$}\cdot
{\bf {\cal H}}
$ causes the hyperfine splitting
structure of atomic levels and will not be discussed here.
The total momentum of the system is given by
\begin{eqnarray}
{\bf P}={\bf P}_{n}+{\bf P}_{e}+{\bf P}_{f}\,,
\end{eqnarray}
where ${\bf P}_{e}=\int d{\bf x}
\psi^{\dag}({\bf x})(-i
\nabla_{\bf x})
\psi({\bf x})$ is the electron-positron field momentum
 and ${\bf P}_{f}=\int d{\bf x}(
{\bf {\cal E}}_{t}({\bf x})\times{\bf {\cal H}}({\bf x}))$
is the electromagnetic field momentum.
Since the total momentum is an integral of the motion
we can restrict our consideration to the center-of-mass
system (${\bf P}=0$) and, so, can  express the nuclear momentum 
in terms of the electron-positron and electromagnetic field
momenta
\begin{eqnarray}
{\bf P}_{n}=-{\bf P}_{e}-{\bf P}_{f}=
-\int d{\bf x}
\psi^{\dag}({\bf x})(-i
\nabla_{\bf x})
\psi({\bf x}) -\int d{\bf x}(
{\bf {\cal E}}_{t}({\bf x})\times{\bf {\cal H}}({\bf x}))\,.
\end{eqnarray}
Using this equation and the translation invariance
we find
\begin{eqnarray}
H&=&\int d{\bf x} \psi^{\dag}({\bf x})
[\mbox{\boldmath $\alpha$}\cdot
(-i\nabla_{\bf x}
-e{\bf A}({\bf x}))+\beta m]\psi({\bf x})\nonumber\\
 &&+\frac{e^{2}}{8\pi}\int d{\bf x}d{\bf y}\frac{
\rho_{e}({\bf x})
\rho_{e}({\bf y})}{|{\bf x}-{\bf y}|}+\frac{1}{2}
\int d{\bf x}
({\bf {\cal E}}_{t}^{2}({\bf x})+{\bf {\cal H}}^{2}({\bf x}))
\nonumber\\
 &&+\frac{e|e|Z}{4\pi}\int d{\bf x}\frac{
\rho_{e}({\bf x})}{|{\bf x}|}
+\frac{1}{2M}
\Biggl[-\int d{\bf x}
\psi^{\dag}({\bf x})(-i
\nabla_{\bf x})
\psi({\bf x})\nonumber\\
&&-\int d{\bf x}(
{\bf {\cal E}}_{t}({\bf x})\times {\bf {\cal H}}({\bf x}))
-|e|Z{\bf A}(0)\Biggr]^{2}\,.
\end{eqnarray}
Here we have omitted the hyperfine interaction term.
The sum of the first four terms in (12)
is the standard Hamiltonian of the electron-positron field
interacting with the quantized electromagnetic field and
with the classical Coulomb field of the nucleus
$V_{c}=-\frac{\alpha Z}{r}$ ( a finite nuclear charge distribution
can be taken into account by replacing $V_{c}$ with the potential
of an extended nucleus (see Sec. 5)). The last term in (12) defines the
nuclear recoil corrections of the first order in $m/M$.
The part of this term containing the electromagnetic field momentum
( ${\bf P}_{f}=\int d{\bf x}\,
({\bf {\cal E}}_{\
t}({\bf x})\times{\bf {\cal H}}({\bf x}))$)
will contribute only in the first and higher orders
in $\alpha$ and, so, will not be discussed here.
It follows, to the zeroth order in $\alpha$ the nuclear
recoil corrections can be calculated by adding to the
standard Hamiltonian the following term
\begin{eqnarray}
H_{M}&=&
\frac{1}{2M}
\int d{\bf x}
\psi^{\dag}({\bf x})(-i
\nabla_{\bf x})
\psi({\bf x})
\int d{\bf y}
\psi^{\dag}({\bf y})(-i
\nabla_{\bf y})
\psi({\bf y})\nonumber\\
&&-\frac{eZ}{M}
\int d{\bf x}
\psi^{\dag}({\bf x})(-i
\nabla_{\bf x})
\psi({\bf x}){\bf A}(0)+\frac{e^{2}Z^{2}}{2M}{\bf A}^{2}(0)\,.
\end{eqnarray}

As is known, for description of
 an atomic system within QED it is convenient
to use the interaction representation in the Furry picture.
In such a theory the normal ordered form of $H_{M}$ taken in
the interaction representation must be added to the interaction
Hamiltonian. A simple way to derive formal expressions for
the energy level shifts consists in using the technique
developed in [14,15].
According to this technique, the energy shift
 $ \Delta E_{a} = E_a-E_a^{(0)}$
 of a single
level $a$ of an $N$-electron atom
is given by the formula
\begin{eqnarray}
\Delta E_a = \frac
{\frac1{2\pi i} \oint_{\Gamma} dE ( E-E_a^{(0)} ) \Delta g_{a}(E)}
{1+ \frac1{2\pi i} \oint_{\Gamma} dE \Delta g_{a}(E)}\,,
\end{eqnarray}
where the contour 
 $\Gamma$  surrounds only the unperturbed level $E =
E_a^{(0)} $ (the contour is assumed to be traversed
counterclockwise), $\Delta g_{a}(E)=g_{a}(E)-g_{a}^{(0)}(E) $,
 $g_{a}(E) $ is defined by 
\begin{eqnarray}
g_{a}(E) \delta (E-E^{\prime}) &=& \frac{2\pi}{i} \frac1{N!}
\int_{-\infty}^{\infty} dp^0_1 \dots dp^0_N \, {dp^{\prime}}^0_1 \dots
{dp^{\prime}}^0_N  \nonumber \\
&& \times \delta (E- p^0_1 - \dots - p^0_N)
 \delta (E^{\prime}- {p^{\prime}}^0_1 - \dots - {p^{\prime}}^0_N)
\nonumber \\
&& \times \langle u_a|G({p^{\prime}}^0_1, \dots
,{p^{\prime}}^0_N;  p^0_1, \dots ,p^0_N) {\gamma}_1^0
\dots {\gamma}_N^0|u_a \rangle
\end{eqnarray}
with $G$ is the usual $N$-electron Green function
\begin{eqnarray}
\lefteqn{
G({p^{\prime}}^0_1,\dots ,{p^{\prime}}^0_N; p^0_1,\dots ,p^0_N)}\nonumber\\
&&={(2\pi )}^{-2N}  \int_{-\infty}^{\infty} dx^0_1 \dots dx^0_N \,
{dx^{\prime}}^0_1 \dots {dx^{\prime}}^0_N \nonumber \\
&&\times \exp{(i{p^{\prime}}^0_1{x^{\prime}}^0_1+ \dots +
i{p^{\prime}}^0_N{x^{\prime}}^0_N - i{p}^0_1{x}^0_1- \dots -
i{p}^0_N{x}^0_N)} \nonumber \\
&&\times \langle 0| T \psi(x_1^{\prime}) \dots \psi(x_N^{\prime})
\overline{\psi}(x_N) \dots \overline{\psi}(x_1) |0 \rangle \,,
\end{eqnarray}
$ \psi (x) $ is the electron-positron field operator in
the Heisenberg representation, $\overline{\psi}(x)=\psi^{\dag}(x)\gamma^{0}$,
$ u_a $ is the unperturbed
atomic wave function, and $g_{a}^{(0)}(E)={(E-E_a^{(0)})}^{-1} $
is the zeroth approximation of $g_{a}(E)$.
In the first order of the perturbation theory we have
\begin{eqnarray}
\Delta E_a^{(1)} = \frac1{2\pi i} \oint_{\Gamma} dE \, (E-E_{a}^{(0)}) \,
  \Delta g_{a}^{(1)}(E) \,.
\end{eqnarray}
The Green function $G$ is constructed using the Wick theorem
after the transition in (16) to the interaction representation.
The diagram technique rules for $G$ are considered in detail in [15].
Including $H_{M}$ in the interaction Hamiltonian gives the following
additional lines and vertices to the diagram technique rules for $G$.
\begin{enumerate}
\item Coulomb contribution.\\

An additional line ("Coulomb-recoil" line) appears to be
\newline\\
\setlength{\unitlength}{0.7mm}
\begin{picture}(60,5)(0,0)
  \multiput(15,2)(2,0){15}{\circle*{1} }
  \put(15,2){\circle*{2}}
  \put(45,2){\circle*{2}}
  \put(30,6){$\omega$}
  \put(12,-6){${\bf x}$}
  \put(45,-6){${\bf y}$}
  \label{intphotline}
\end{picture}
          $ \frac{i}{2\pi} \frac{\delta_{kl}}{M}
\int_{-\infty}^{\infty}d\omega\,. $ \\
\newline
This line joins two vertices each of which corresponds to
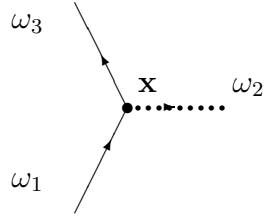
\newline\\
\begin{picture}(60,60)(0,0)
  \put(50,30){\line(-1,2){10}}
  \put(50,30){\line(-1,-2){10}}
  \multiput(50,30)(2,0){10}{\circle*{1}}
  \put(50,30){\circle*{2}}
  \put(52,33){{\bf x}}
  \put(70,33){$\omega_{2}$}
  \put(28,15){$\omega_{1}$}
  \put(28,45){$\omega_{3}$}
  \put(58,30){\vector(1,0){1}}
  \put(46,38){\vector(-1,2){1}}
  \put(46,22){\vector(1,2){1}}
\label{vertex}
\end{picture}  
       $ -i 2\pi\gamma_{0}\delta(\omega_{1}-\omega_{2}-\omega_{3})
            \int d{\bf x}\,p_{k} \;,$ \newline \\ 
\newline
where
 ${\bf p}=
-i\nabla_{\bf x}$ and $k=1,2,3$.

\item One-transverse-photon contribution.\\

An additional vertex on an electron line appears to be
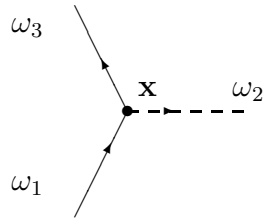
\newline
\begin{picture}(60,60)(0,0)
  \put(50,30){\line(-1,2){10}}
  \put(50,30){\line(-1,-2){10}}
  \multiput(50,30)(4,0){6}{\line(1,0){2}}
  \put(50,30){\circle*{2}}
  \put(52,33){{\bf x}}
  \put(70,33){$\omega_{2}$}
  \put(28,15){$\omega_{1}$}
  \put(28,45){$\omega_{3}$}
  \put(58,30){\vector(1,0){1}}
  \put(46,38){\vector(-1,2){1}}
  \put(46,22){\vector(1,2){1}}
\label{vertex1}
\end{picture}  
   $ -i 2\pi\gamma_{0}
\delta(\omega_{1}-\omega_{2}-\omega_{3})
       \frac{eZ}{M} \int d{\bf x}\,p_{k} \;,$ \newline \\ 

The transverse photon line attached to this vertex (at
the point ${\bf x}$) is
\newline\\
\begin{picture}(60,5)(0,0)
  \multiput(15,2)(4,0){9}{\line(1,0){2} }
  \put(15,2){\circle*{2}}
  \put(30,6){$\omega$}
  \put(12,-6){${\bf x}$}
  \put(47,-6){${\bf y}$}
  \label{intphotline1}
\end{picture}
          $ \frac{i}{2\pi}
\int_{-\infty}^{\infty}d\omega D_{kl}(\omega,{\bf y})\,. $ \\
\newline\\
At the point ${\bf y}$ this line is to be attached to an usual
vertex in which we have $-ie\gamma_{0}\alpha_{l}2\pi
\delta(\omega_{1}-\omega_{2}-\omega_{3})\int d{\bf y}$ (see [15]),
where $\alpha_{l}$ ($l=1,2,3$)
are the usual Dirac matrices (we note here
that in the notations of [15]
 $\alpha^{\mu}=
(1,\mbox{\boldmath$\alpha$})$ and      
 $\alpha_{\mu}=
(1,-\mbox{\boldmath$\alpha$})$ ).

\item Two-transverse-photon contribution.\\

An additional line ("two-transverse-photon-recoil" line)
appears to be
\newline\\
\begin{picture}(60,5)(0,0)
  \multiput(15,2)(4,0){9}{\line(1,0){2} }
  \put(32,2){\circle*{2}}
  \put(12,-6){${\bf x}$}
  \put(47,-6){${\bf y}$}
  \put(30,6){$\omega$}
  \label{intphotline1}
\end{picture}
          $ \frac{i}{2\pi}\frac{e^{2}Z^{2}}{M}
\int_{-\infty}^{\infty}d\omega
 D_{il}(\omega,{\bf x})
 D_{lk}(\omega,{\bf y})\,. $ 
\newline\\
This line joins  usual vertices (see the previous item).
\end{enumerate}

An important advantage of the approach considered here,
in comparison with one developed in [1,11], consists
in that the present method is suitable for arbitrary local
potential $V(r)$ (e.g., a local version of the Hartree-Fock 
potential) used as the zero approximation. In addition,
the transition to the formalism in which the role of the vacuum
is played by closed shells can simply be realized by changing
the sign of $i0$ in the electron propagator denominators
corresponding to the closed shells.

\section{One electron over closed shells}

Let us consider an atom with one electron over closed shells.
In the zero approximation the electrons of the atom interact
with the potential $V(r)$ which can be chosen to include
approximately the electron-electron interaction. In the
formalism with the closed shell states as well as the negative
energy states refered to the vacuum, the electron propagator
is given by
\begin{eqnarray}
S(\omega,{\bf x},{\bf y})=\sum_{n}\frac{\psi_{n}({\bf x})
\overline{\psi}_{n}({\bf y})}{\omega-\varepsilon_{n}+i\eta_{n}0}\,,
\end{eqnarray}
where $\eta_{n}=\varepsilon_{n}-\varepsilon_{F}$ and $\varepsilon_{F}$
is the Fermi energy which is chosen to be higher than the
 one-electron closed
shell energies and lower than the energies of
the one-electron
states over the closed shells. In the simplest case of
an one-electron atom $\eta_{n}=\varepsilon_{n}$.

To find the Coulomb nuclear recoil correction we have to calculate 
the contribution of the diagram shown in Fig. 1. According to the
diagrams technique rules given in the previous section and [15]
we obtain
\begin{eqnarray}
\Delta g_{a}^{(1)}=\frac{1}{(E-E_{a}^{(0)})^{2}}\frac{1}{M}\frac{i}{2\pi}
\int_{-\infty}^{\infty}d\omega \sum_{n}\frac
{\langle a| p_{i}|n\rangle\langle n|p_{i}|a\rangle}
{\omega-\varepsilon_{n}+i\eta_{n}0}\,.
\end{eqnarray}
The formula (17) gives
\begin{eqnarray}
\Delta E_{c}=\frac{1}{M}\frac{i}{2\pi}
\int_{-\infty}^{\infty}d\omega \sum_{n}\frac
{\langle a| p_{i}|n\rangle\langle n|p_{i}|a\rangle}
{\omega-\varepsilon_{n}+i\eta_{n}0}\,.
\end{eqnarray}
Using the identities
\begin{eqnarray}
\frac{1}{x+i0}&=&\frac{\pi}{i}\delta(x)+{\rm P}\frac{1}{x}\,,\\
\frac{1}{x-i0}&=&\pi i\delta(x)+{\rm P}\frac{1}{x}
\end{eqnarray}
one can get
\begin{eqnarray}
\Delta E_{c}&=&
\frac{1}{2M}
 \sum_{n} \frac{\eta_{n}}{|\eta_{n}|}
|\langle a|{\bf p}|n\rangle|^{2}\nonumber\\
&=&\frac{1}{2M}
 \langle a|{\bf p}^{2}|a\rangle
-\frac{1}{M}
 \sum_{\varepsilon_{n}<\varepsilon_{F}} 
|\langle a|{\bf p}|n\rangle|^{2}\,.
\end{eqnarray}
The one-transverse-photon nuclear recoil correction corresponds
to the diagrams shown in Fig. 2. A similar calculation gives
\begin{eqnarray}
\Delta E_{tr(1)}&=&\frac{4\pi\alpha Z}{M}\frac{i}{2\pi}
\int_{-\infty}^{\infty}d\omega\,\sum_{n}\Biggl\{
\frac{
\langle a|p_{i}|n\rangle\langle n|
\alpha_{k} D_{ik}(\varepsilon_{a}-\omega)|a\rangle}
{\omega-\varepsilon_{n}+i\eta_{n}0}
\nonumber\\
&&
+\frac{
\langle a|
\alpha_{k} D_{ik}(\varepsilon_{a}-\omega)|n\rangle
\langle n|p_{i}|a\rangle}
{\omega-\varepsilon_{n}+i\eta_{n}0}\Biggr\}\,.
\end{eqnarray}
By using the identity
$$
\frac{1}{\omega-\varepsilon_{n}+i\eta_{n}0}
=
\frac{1}{\omega-\varepsilon_{a}+i0}
+\frac{\varepsilon_{n}-\varepsilon_{a}}
{(\omega-\varepsilon_{a}+i0)
(\omega-\varepsilon_{n}+i\eta_{n}0)}
$$
and the equation (21), the expression (24) can easily
be transformed to the following
\begin{eqnarray}
\Delta E_{tr(1)}&=&\Delta E_{tr(1)}^{(1)}+
\Delta E_{tr(1)}^{(2)}\,,\nonumber\\
\Delta E_{tr(1)}^{(1)}&=&
\frac{4\pi\alpha Z}{2M}
\langle a|(p_{i}
\alpha_{k} D_{ik}(0)
+\alpha_{k} D_{ik}(0)p_{i})|a\rangle\,,\\
\Delta E_{tr(1)}^{(2)}&=&\frac{4\pi\alpha Z}{M}
\int_{-\infty}^{\infty}d\omega\,\delta_{+}(\omega-\varepsilon_{a})\,
\sum_{n}\Biggl\{
\frac{
\langle a|[p_{i},V]|n\rangle\langle n|
\alpha_{k} D_{ik}(\varepsilon_{a}-\omega)|a\rangle}
{\omega-\varepsilon_{n}+i\eta_{n}0}
\nonumber\\
&&
-\frac{
\langle a|
\alpha_{k} D_{ik}(\varepsilon_{a}-\omega)|n\rangle
\langle n|[p_{i},V]|a\rangle}
{\omega-\varepsilon_{n}+i\eta_{n}0}\Biggr\}\,.
\end{eqnarray}
The two-transverse-photon nuclear recoil correction is defined
by the diagram shown in Fig. 3. We find
\begin{eqnarray}
\Delta E_{tr(2)}&=&\frac{(4\pi\alpha Z)^{2}}{M}\frac{i}{2\pi}
\int_{-\infty}^{\infty}d\omega\,\sum_{n}\nonumber\\
&&\times\frac{
\langle a|\alpha_{i}D_{il}(\varepsilon_{a}-\omega)|n\rangle\langle n|
\alpha_{k} D_{lk}(\varepsilon_{a}-\omega)|a\rangle}
{\omega-\varepsilon_{n}+i\eta_{n}0}\,.
\end{eqnarray}
As it follows from the equations (20), (24), and (27),
the sum of all the contributions can be written
in the following compact form
\begin{eqnarray}
\Delta E_{tot}&=&\frac{1}{M}\frac{i}{2\pi}\int_{-\infty}^{\infty}d\omega\,
\langle a|(p_{i} +4\pi\alpha Z\alpha_{l} D_{li}(\omega))\nonumber\\
&&\times G(\omega +\varepsilon_{a})
( p_{i} +4\pi\alpha Z\alpha_{m} D_{mi}(\omega))|a\rangle\,,
\end{eqnarray}
where $G(\varepsilon)=\sum_{n}\frac{|n\rangle\langle n|}{\varepsilon
-\varepsilon_{n}+i\eta_{n} 0}$ is the electron Green function.

In the case of a hydrogenlike atom, the expressions derived
here coincide with  ones
given in Sec. 1.

\section{Two electrons over closed shells}

Consider now an atom with two electrons over closed shells
(a general case of $N$ electrons over closed shells can be considered
in the same way). For simplicity, we take as the unperturbed
wave function the one-determinant wave function
  \begin{equation}
   u=\frac{1}{\sqrt{2}}\sum_{P}(-1)^{P}\psi_{P a}({\bf x_{1}})
        \psi_{P b}({\bf x_{2}})\,.
\end{equation}
The nuclear recoil correction is the sum of the one-electron
and two-electron contributions. Using the diagram technique rules
from [15] and the Sec. 2 and the formula (17) one easily
finds that the one-electron contribution is equal to the sum
of the expressions (28) for $a$ and $b$ states. The two-electron
contributions correspond to the diagrams shown in Fig. 4-6.
The two-electron Coulomb contribution is
\begin{eqnarray}
 \Delta E_{c}^{(int)}&=&
  \frac{1}{M} 
    \frac{1}{2\pi i}\oint_{\Gamma} dE (E-E^{(0)})
\Biggl\{  \Bigl(\frac{i}{2\pi}\Bigr)^{2}
        \int_{-\infty}^{\infty}dp^{0}dp^{\prime 0}\nonumber\\
&&\times    \sum_{P}(-1)^{P}
\frac{1}{p^{\prime 0}-\varepsilon_{P a}+i0}\;\;  
           \frac{1}{E-p^{\prime 0}-\varepsilon_{P b}+i0}\nonumber\\
&&\times     \frac{1}{p^{0}-\varepsilon_{a}+i0}\;\;
               \frac{1}{E-p^{0}-\varepsilon_{b}+i0}\nonumber\\ 
&&\times  
\langle Pa|p_{i}|a\rangle
\langle Pb|p_{i}|b\rangle\Biggr\}\,.
\end{eqnarray}
Integrating over $p^{0}$, $p^{\prime 0}$, and $E$ we get
\begin{eqnarray}
\Delta E_{c}^{(int)}=
\frac{1}{M} \sum_{P}(-1)^{P}
\langle Pa|p_{i}|a\rangle
\langle Pb|p_{i}|b\rangle\,.
\end{eqnarray}
A similar calculation of the one-transverse-photon contribution
gives
\begin{eqnarray}
\Delta E_{tr(1)}^{(int)}&=&
\frac{4\pi\alpha Z}{M} \sum_{P}(-1)^{P}
\Bigl[\langle Pa|p_{i}|a\rangle
\langle Pb|\alpha_{k}D_{ki}
(\varepsilon_{Pb}-\varepsilon_{b})
|b\rangle\nonumber\\
&&+\langle Pa|\alpha_{k}D_{ki}
(\varepsilon_{Pa}-\varepsilon_{a})
|a\rangle
\langle Pb|p_{i}|b\rangle\Bigr]\,.
\end{eqnarray}
Finally, for the two-transverse-photon contribution we find
\begin{eqnarray}
\Delta E_{tr(2)}^{(int)}&=&
\frac{(4\pi\alpha Z)^{2}}{M} \sum_{P}(-1)^{P}
[\langle Pa|\alpha_{k}D_{ki}
(\varepsilon_{Pa}-\varepsilon_{a})
|a\rangle\nonumber\\
&&\times\langle Pb|\alpha_{m}D_{mi}
(\varepsilon_{Pb}-\varepsilon_{b})
|b\rangle\,.
\end{eqnarray}
The sum of the two-electron contributions (31)-(33)
 can be written in the following
compact form 
\begin{eqnarray}
\Delta E_{tot}^{(int)}&=&
\frac{1}{M} \sum_{P}(-1)^{P}
\langle Pa|p_{i}+4\pi\alpha Z\alpha_{l}D_{li}
(\varepsilon_{Pa}-\varepsilon_{a})
|a\rangle\nonumber\\
&&\times\langle Pb|p_{i}+4\pi\alpha Z\alpha_{m}D_{mi}
(\varepsilon_{Pb}-\varepsilon_{b})
|b\rangle\,.
\end{eqnarray}

The formulas (31)-(34) coincide with the related expressions 
found for high-$Z$ few-electron atoms in [11] (see also [7]).
The difference is only the present expressions (31)-(34)
are not restricted to the case of the pure Coulomb zero approximation.

\section{Composite nuclear structure}

The problem of the composite nuclear structure in the nuclear recoil
theory was first discussed by Salpeter [6]. In Ref. [6], it was shown
that the calculations based on  the assumption that the nucleus
is a point Dirac particle of electric charge $|e|Z$ and  mass $M$
are valid for composite nuclei (independently of the nuclear spin),
if the distance between the nuclear levels is large compared
with the distance between the atomic (electrons + field)
levels contributing to the nuclear recoil effect.
In this section we consider how this result can be derived within
the approach developed in the present paper.

Let us assume, for simplicity, that the nucleus is a bound state
of a two-particle system (e.g., a core with a mass $m_{1}$ and a charge
$e_{1}$ and a valent nucleon with a mass $m_{2}$ and a charge $e_{2}$).
In this case the sum of the last three terms in equation (9)
must be replaced by $H_{1}+H_{2}+H_{3}$, where
\begin{eqnarray}
H_{1}&=&\frac{e}{4\pi}\int d{\bf x}
\rho_{e}({\bf x}) \Biggl(
\frac{e_{1}}{|{\bf x}-{\bf x}_{1}|}
+\frac{e_{2}}{|{\bf x}-{\bf x}_{2}|}\Biggr)\,,\\
H_{2}&=&
\frac{1}{2m_{1}}
({\bf p}_{1}-e_{1}{\bf A}({\bf x}_{1}))^{2}
+\frac{1}{2m_{2}}
({\bf p}_{2}-e_{2}{\bf A}({\bf x}_{2}))^{2}\nonumber\\
&&+U({\bf x}_{1}-{\bf x}_{2})\,,\\
H_{3}&=&
-\mbox{\boldmath $\mu$}_{s}^{(1)}\cdot
{\bf {\cal H}}({\bf x}_{1})
-\mbox{\boldmath $\mu$}_{s}^{(2)}\cdot
{\bf {\cal H}}({\bf x}_{2})\,.
\end{eqnarray}
Here $U({\bf x}_{1}-{\bf x}_{2})$ describes the interaction
between the nuclear particles
(for simplicity, we assume that $U$ does not depend of the spins)
 and $
\mbox{\boldmath $\mu$}_{s}^{(1)}$ and
$\mbox{\boldmath $\mu$}_{s}^{(2)}$ are the intristic magnetic
moments of the nuclear particles. Introducing the center-of-nucleus-mass
variables
\begin{eqnarray}
{\bf X}_{n}=\frac{
m_{1}{\bf x}_{1}
+m_{2}{\bf x}_{2}}{m_{1}+m_{2}}\,,\;\;\;\;\;\;
{\bf x}_{n}={\bf x}_{1}-{\bf x}_{2}\,,
\end{eqnarray}
we have
\begin{eqnarray}
{\bf p}_{1}=\frac{m_{1}}{m_{1}+m_{2}}{\bf P}_{n}+{\bf p}_{n}
\,,\;\;\;\;\;\;
{\bf p}_{2}=\frac{m_{2}}{m_{1}+m_{2}}{\bf P}_{n}-{\bf p}_{n}\,,
\end{eqnarray}
where
 ${\bf P}_{n}=
-\nabla_{{\bf X}_{n}}$ and
 ${\bf p}_{n}=
-\nabla_{{\bf x}_{n}}$ .
As in the Sec. 2, we can restrict our consideration to
the center-of-atom-mass system (
${\bf P}= {\bf P}_{n}+ {\bf P}_{e}+{\bf P}_{f}=0$).
So, the total nuclear momentum ${\bf P}_{n}$ is given
by the equation (11). In terms of the variables ${\bf X}_{n}$
and ${\bf x}_{n}$ the operator $H_{2}$ can be represented
in the form
\begin{eqnarray}
H_{2}=H_{\mu}+H_{M}\,,
\end{eqnarray}
where 
\begin{eqnarray}
H_{\mu}=\frac{{\bf p}_{n}^{2}}{2\mu}+U({\bf x}_{n})\,,
\end{eqnarray}
and $\mu=(m_{1}m_{2})/(m_{1}+m_{2})$. The Hamiltonian
$H_{\mu}$ describes the intristic states of the nucleus.
Let us denote the wave function of the nuclear state which is
 under the consideration
by
 $\phi_{a}({\bf x}_{n})$.
 The wave function of the whole 
system in the zero approximation 
is the product of $\phi_{a}({\bf x}_{n})$ and the atomic
wave function calculated using the operator
$\langle\phi_{a}|H_{1}|\phi_{a}\rangle$ 
as the interaction
with the nucleus (we assume here and subsequently that the  distance
between the nuclear energy levels is large compared with
the distance between the atomic levels).
The operator $\langle\phi_{a}|H_{1}|\phi_{a}\rangle$ 
describes the interaction of electrons with the extended
nucleus charge. The $m/M$ corrections are calculated by
perturbation theory. 
Using the fact that the nuclear size ($\sim|{\bf x}_{n}|$)
is much smaller than the atomic size, we expand the vector
${\bf A}$ in powers of ${\bf x}_{n}$. Taking into account
that  $\phi_{a}({\bf x}_{n})$ is of a definite parity we find
to the lowest orders 
\begin{eqnarray}
\lefteqn{
\langle\phi_{a}|H_{M}
|\phi_{a}\rangle}\nonumber\\
&= 
\langle\phi_{a}|\Biggl\{\frac{{\bf P}_{n}^{2}}{2M}
-\frac{(e_{1}+e_{2})}{2M}(
{\bf P}_{n}\cdot
 {\bf A}({\bf X}_{n})
+ {\bf A}({\bf X}_{n})\cdot
{\bf P}_{n})\nonumber\\
&+
\Bigl(
\frac{e_{1}^{2}}{2m_{1}}+
\frac{e_{2}^{2}}{2m_{2}}\Bigr)
{\bf A}^{2}({\bf X}_{n})
-\frac{1}{2M}
\Bigl(
\frac{e_{1}m_{2}}{m_{1}}+
\frac{e_{2}m_{1}}{m_{2}}\Bigr)\nonumber\\
&\times
[{\bf p}_{n}
({\bf x}_{n}\cdot
\nabla_{{\bf X}_{n}})
{\bf A}({\bf X}_{n})
+({\bf x}_{n}\cdot
\nabla_{{\bf X}_{n}})
{\bf A}({\bf X}_{n}){\bf p}_{n}]\Biggr\}
|\phi_{a}\rangle\,.
\end{eqnarray}
The last term in (42) can be transformed to 
\begin{eqnarray}
&-\frac{1}{2M}\Bigl(
\frac{e_{1}m_{2}}{m_{1}}+
\frac{e_{2}m_{1}}{m_{2}}\Bigr)\int d{\bf x}_{n}
\phi_{a}^{*}({\bf x}_{n})
({\bf x}_{n}\times {\bf p}_{n})
\phi_{a}({\bf x}_{n})
\,{\bf{\cal H}}({\bf X}_{n})\nonumber\\
&=
-\Bigl(
\frac{e_{1}}{2m_{1}}
\langle{\bf l}_{1}\rangle
+\frac{e_{2}}{2m_{2}}
\langle{\bf l}_{2}\rangle\Bigr)
\,{\bf{\cal H}}({\bf X}_{n})\,,
\end{eqnarray}
where ${\bf l}_{1}$ and
 ${\bf l}_{2}$ are the orbital
moments of the nuclear particles in the center-of-nucleus-mass system.
 Adding this term to the
term $H_{3}$ gives the total operator of the hyperfine interaction
$-\mbox{\boldmath $\mu$}\cdot
{\bf {\cal H}}({\bf X}_{n})$, where
$\mbox{\boldmath $\mu$}=
\mbox{\boldmath $\mu$}_{l}^{(1)}+\mbox{\boldmath $\mu$}_{s}^{(1)}
+\mbox{\boldmath $\mu$}_{l}^{(2)}+\mbox{\boldmath $\mu$}_{s}^{(2)}$
is the total magnetic moment of the nucleus and
$\mbox{\boldmath $\mu$}_{l}^{(i)}=\frac{e_{i}}{2m_{i}}
{\bf l}_{i}$. Due to the operator $H_{M}$ contains
the term
\begin{eqnarray}
H_{M}'=
-\Bigl(
\frac{e_{1}}{m_{1}}-
\frac{e_{2}}{m_{2}}\Bigr){\bf p}_{n}\cdot {\bf A}({\bf X}_{n})\,,
\end{eqnarray}
there is a contribution of the order $m/M$ from the second
order of the perturbation theory. For a state $a$ of the
whole system we have
\begin{eqnarray}
\Delta E_{a}'=\Bigl(
\frac{e_{1}}{m_{1}}-
\frac{e_{2}}{m_{2}}\Bigr)^{2}
\sum_{n\not=a}\frac{
\langle a|
{\bf p}_{n}\cdot {\bf A}({\bf X}_{n})|n\rangle
\langle n|
{\bf p}_{n}\cdot {\bf A}({\bf X}_{n})|a\rangle}
{E_{a}-E_{n}}
\,.
\end{eqnarray}
Assuming that the energy difference between the nuclear state which
is under the consideration and the other nuclear states
contributing to the sum in the equation (45) is large compared with
the corresponding energy differences between the atomic
(electrons + field) states which give a dominant contribution
to $\Delta E_{a}'$, we replace $E_{a}-E_{n}$ in Eq. (45) with
$\epsilon_{a}-\epsilon_{n}$, where 
$\epsilon_{a}$ and $\epsilon_{n}$ are the nuclear energies.
Using the identity ${\bf p}_{n}=i\mu[H_{\mu},{\bf x}_{n}]$
we find 
\begin{eqnarray}
\Delta E_{a}'&=&
\Bigl(
\frac{e_{1}}{m_{1}}-
\frac{e_{2}}{m_{2}}\Bigr)^{2}\langle \Phi_{a}|{\bf A}^{2}({\bf X}_{n})
|\Phi_{a}\rangle
\frac{i}{2}\mu
\sum_{n\not=a}\frac{1}{\epsilon_{a}-\epsilon_{n}}\nonumber\\
&&\times
[\langle\phi_{ a}|[H_{\mu},
{\bf x}_{n}]|\phi_{n}\rangle
\langle\phi_{ n}|
{\bf p}_{n}
|\phi_{a}\rangle
+
\langle\phi_{ a}|
{\bf p}_{n}
|\phi_{n}\rangle
\langle\phi_{ n}|[H_{\mu},
{\bf x}_{n}]|\phi_{a}\rangle]\nonumber\\
&&=\Bigl(
\frac{e_{1}}{m_{1}}-
\frac{e_{2}}{m_{2}}\Bigr)^{2}\langle \Phi_{a}|{\bf A}^{2}({\bf X}_{n})
|\Phi_{a}\rangle
\frac{i}{2}\mu
\sum_{n\not=a}\nonumber\\
&&\times
[\langle\phi_{ a}|
{\bf x}_{n}|\phi_{n}\rangle
\langle\phi_{ n}|
{\bf p}_{n}
|\phi_{a}\rangle
-
\langle\phi_{ a}|
{\bf p}_{n}
|\phi_{n}\rangle
\langle\phi_{ n}|
{\bf x}_{n}|\phi_{a}\rangle]\nonumber\\
&&=
\Bigl(
\frac{e_{1}}{m_{1}}-
\frac{e_{2}}{m_{2}}\Bigr)^{2}\langle \Phi_{a}|{\bf A}^{2}({\bf X}_{n})
|\Phi_{a}\rangle
\frac{i}{2}\mu
\langle\phi_{ a}|[{\bf x}_{n},{\bf p}_{n}]
|\phi_{a}\rangle\nonumber\\
&&=
-\frac{1}{2}\Bigl(
\frac{e_{1}}{m_{1}}-
\frac{e_{2}}{m_{2}}\Bigr)^{2}\mu
\langle \Phi_{a}|{\bf A}^{2}({\bf X}_{n})
|\Phi_{a}\rangle
\,,
\end{eqnarray}
where $|\phi_{a}\rangle$ is the nuclear wave function and
$|\Phi_{a}\rangle$ is the atomic (electrons + field)
wave function. Combaining this term with the equation (42)
we find that the nuclear recoil correction of the first order
in $m/M$ is defined by the operator
\begin{eqnarray}
H_{M}&=&
\frac{{\bf P}_{n}^{2}}{2M}
-\frac{(e_{1}+e_{2})}{2M}(
{\bf P}_{n}\cdot
 {\bf A}({\bf X}_{n})
+ {\bf A}({\bf X}_{n})\cdot
{\bf P}_{n})\nonumber\\
&&+\frac{(e_{1}+e_{2})^{2}}{2M}
{\bf A}^{2}({\bf X}_{n})\nonumber\\
&&= \frac{1}{2M}
({\bf P}_{n}-|e|Z
 {\bf A}({\bf X}_{n}))^{2}\,.
\end{eqnarray}
So, within the approximations made in the derivation of this
result, the effects of the composite nuclear structure
can be neglected.

\section*{Acknowledgements}
Valuable conversations with I.P.Grant, S.G.Karshenboim, I.B.Khriplovich,
G.Soff, D.A.Tel'nov, and A.S.Yelkhovsky are gratefully acknowledged.
 The research described in this
publication was made possible in part by  Grant No. 95-02-05571a from
RFBR.

\newpage
\begin{figure}
\setlength{\unitlength}{1mm}
\begin{center}
\begin{picture}(100,60)(20,0)
   \put(40,10){\line(0,1){30}}
   \multiput(40,17)(-2,2){5}{\circle*{1}}
   \multiput(40,33)(-2,-2){5}{\circle*{1}}
   \put(40,33){\circle*{2}}
   \put(40,17){\circle*{2}}
   \end{picture}
\caption{ Coulomb nuclear recoil diagram.}
\end{center}
\end{figure}
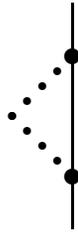

\begin{figure}
\setlength{\unitlength}{1mm}
\begin{center}
\begin{picture}(100,60)(20,0)
   \put(20,10){\line(0,1){30}}
  \put(60,10){\line(0,1){30}}
   \multiput(20,17)(-2,2){5}{\line(-1,0){1}}
   \multiput(20,33)(-2,-2){5}{\line(-1,0){1}}
   \multiput(60,17)(2,2){5}{\line(1,0){1}}
   \multiput(60,33)(2,-2){5}{\line(1,0){1}}
   \put(60,33){\circle*{2}}
   \put(20,17){\circle*{2}}
   \put(18,1){a}
   \put(58,1){b} 
   \end{picture}
\caption{ One-transverse-photon nuclear recoil diagrams.}
\end{center}
\end{figure}

\begin{figure}
\setlength{\unitlength}{1mm}
\begin{center}
\begin{picture}(100,60)(20,0)
   \put(40,10){\line(0,1){30}}
   \multiput(40,17)(-2,2){4}{\line(-1,0){1}}
   \multiput(40,33)(-2,-2){4}{\line(-1,0){1}}
   \put(33,25){\circle*{2}}
   \end{picture}
\caption{ Two-transverse-photon nuclear recoil diagram.}
\end{center}
\end{figure}

\begin{figure}
\setlength{\unitlength}{1mm}
\begin{center}
\begin{picture}(100,60)(20,0)
  \put(20,10){\line(0,1){30}}
   \put(40,10){\line(0,1){30}}
   \multiput(20,25)(2,0){10}{\circle*{1}}
   \put(20,25){\circle*{2}}
   \put(40,25){\circle*{2}}
   \end{picture}
\caption{Two-electron Coulomb nuclear recoil diagram.}
\end{center}
\end{figure}
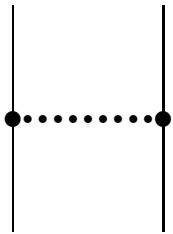
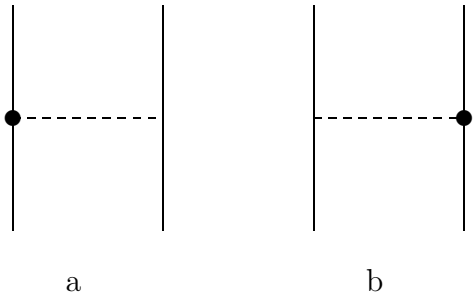
\begin{figure}
\setlength{\unitlength}{1mm}
\begin{center}
\begin{picture}(120,60)(20,0)
  \put(20,10){\line(0,1){30}}
  \put(40,10){\line(0,1){30}}
  \multiput(20,25)(2,0){10}{\line(1,0){1}}
   \put(20,25){\circle*{2}}
   \put(27,2){a}
   \put(67,2){b} 
  \put(60,10){\line(0,1){30}}
  \put(80,10){\line(0,1){30}}
  \multiput(60,25)(2,0){10}{\line(1,0){1}}
   \put(80,25){\circle*{2}}
\end{picture}
\caption{Two-electron one-transverse-photon nuclear recoil diagrams.}
\end{center}
\end{figure}
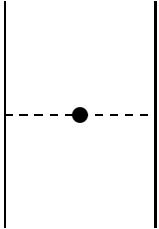
\begin{figure}
\setlength{\unitlength}{1mm}
\begin{center}
\begin{picture}(100,60)(20,0)
  \put(20,10){\line(0,1){30}}
   \put(40,10){\line(0,1){30}}
   \multiput(20,25)(2,0){10}{\line(1,0){1}}
   \put(30,25){\circle*{2}}
   \end{picture}
\caption{Two-electron two-transverse-photon nuclear recoil diagram.}
\end{center}
\end{figure}

\end{document}